# Delay Sensitivity Classification of Cloud Gaming Content

Saeed Shafiee Sabet[1,2], Steven Schmidt[2], Saman Zadtootaghaj[2], Carsten Griwodz[3,] Sebastian Moller[1,4]

[1] SimulaMet, Oslo, Norway
[2] Quality and Usability Lab, TU Berlin, Berlin, Germany
[3] Department of Informatics, University of Oslo, Oslo, Norway
[4] German Research Center for Artificial Intelligence (DFKI), Berlin, Germany

Saeed@simula.no, Griff@ifi.uio.no, Steven.schmidt@tu-berlin.de, saman.zadtootaghaj@qu.tu-berlin.de, Sebastian.moeller@tu-berlin.de

## ABSTRACT

Cloud Gaming is an emerging service that catches growing interest in the research community as well as industry. Cloud Gaming require a highly reliable and low latency network to achieve a satisfying Quality of Experience (QoE) for its users. Using a cloud gaming service with high latency would harm the interaction of the user with the game, leading to a decrease in playing performance and, thus players frustrations. However, the negative effect of delay on gaming QoE depends strongly on the game content. At a certain level of delay, a slow-paced card game is typically not as delay sensitive as a shooting game. For optimal resource allocation and quality estimation, it is highly important for cloud providers, game developers, and network planners to consider the impact of the game content. This paper contributes to a better understanding of the delay impact on QoE for cloud gaming applications by identifying game characteristics influencing the delay perception of the users. In addition, an expert evaluation methodology to quantify these characteristics as well as a delay sensitivity classification based on a decision tree are presented. The results indicated an excellent level of agreement, which demonstrates the reliability of the proposed method. Additionally, the decision tree reached an accuracy of 90% on determining the delay sensitivity classes which were derived from a large dataset of subjective input quality ratings during a series of experiments.

## 1 Introduction

The video gaming market is growing. In 2019, it generated $148.8 billion [1] and was projected to be a $300 billion industry by 2025 [2]. On the other hand, new services such as Cloud Gaming (CG) that can expand the gaming industry even further are emerging. The idea behind CG is to move the heavy processing of rendering the game to a cloud server and stream the rendered scene as a video sequence to the client-side. Using such a service, a user can start playing anytime, anywhere, and on any device. In addition, users do not need to have a console or a PC with a high-end graphic card to play the latest games. CG also provides many benefits to the game industry, such as removing the concerns related to cross-platform development and software piracy.

However, despite these benefits, CG must overcome many challenges, as it requires a stable network with high bandwidth and low latency connection to be deliver the games smoothly to the users and create a good Quality of Experience (QoE). However, depending on the game content these requirements might be different. Various studies have shown that the sensitivity of games toward the delay is strongly content dependent. A game scenario with many dull moments such as in a typical card game is not as sensitive towards the network delay as a shooting game.

Often games from the same genre are similar in respect to their sensitivity towards delay. However, as shown in [3] even within the same game often very different scenarios can exist, e.g., a game like Grand Theft Auto (GTA) consists of a mixture of various genres: some scenarios are similar to a racing game, some similar to a shooting game, card playing game, and etc. Therefore, genre classification is too broad and does not accurately show the game requirement with respect to the delay.

For optimal resource allocation and quality estimation, it is highly important for cloud providers, game developers, and network planners to consider the impact of the game content. Therefore, being aware of the game requirements is necessary for providing a better QoE for their users. This paper first identifies the important characteristics that make a game sensitive to delay and proposes an expert evaluation method in which experienced players should quantify certain characteristics of gaming content to enable a delay sensitivity classification based on a decision tree.

To derive these game characteristics, a focus group using experienced players in gaming was conducted. Nine characteristics potentially influencing the sensitivity of a game towards delay were identified. Afterward, in a second study 14 experienced gamers were invited to quantify these characteristics for 30 different games. In combination with subjective ratings of Input Quality (IQ) assessed in a series of experiments following the ITU-T Rec. P.809, a delay sensitivity classification based on a decision tree, was developed.

The remainder of the paper is organized as follows: Section 2 gives an overview of previous works. Section 3 explains the focus group methodology and identified game characteristics influencing the delay sensitivity of games. Section 4 describes the quantification method and Section 5 describe the development and validation of the decision tree. Section 6 discusses the results. Lastly, a conclusion is drawn in section 7.

## 2 Related Work

The negative influence of delay on gaming QoE has been shown in many papers. Jarschel et al. showed that delay is one of the most influencing factors on the gaming QoE [4]. Beigbeder showed that the delay degrades the overall QoE, and the user's performance [5]. Sabet et al. investigated the relationship between delay, gaming QoE and performance on the combined way in cloud gaming [6].

Furthermore, the game itself was identified as an important influencing factor [7] on gaming QoE in various ways. There have been many efforts to classify games in different areas. Djaouti classified games based on their rules and goals. The authors defined ten game bricks such as move, shoot, avoid and explore [8]. Lee et al. classified games based on their scoring system. The author categorized the games' scoring system based on three characteristics: preservability, controllability, and relationship to achievement [9].

Zadtootaghaj et al. classified games with respect to their video complexity by means of a decision tree [10]. The authors used a focus group to identify characteristics possibly influencing the video complexity and then built a decision tree using these characteristics.

A commonly used classification for marketing purposes is genre classification, which classifies games into genres such as fight, sports, shooter, and racing. Quax et al. [11] showed that games that belong to the same game genre are often similar in respect to their delay sensitivity. However, the genre classification is too broad, and games can be a mixture of different scenarios. Schmidt et al. [3] showed that even within the same game, different scenarios might have different sensitivity toward network delay.

Claypool et al. [12] categorized the games with respect to their delay sensitivity using two game characteristics, precision and deadline. Precision is the degree of accuracy is required to complete a task and the deadline is the available time for the user to complete the task. Furthermore, Eg et al. [13] showed players performance under delay is dependent on the task difficulty and they modeled the influence of delay on target selection using the speed of the target [14]. In addition to precision and speed, Sabet et al. [15] showed the influence of delay is lower in games with predictable actions.

## 3 Identification of Characteristics

To identify characteristics influencing the delay sensitivity, a focus group was conducted. The focus group methodology has gained much popularity in the social sciences to find out the most appropriate research result. It is a qualitative research method that is used to explore person's ways of understanding and experiences. The methodology of the focus group consists of an interview of different people with a moderator. A moderator leads the discussion within the group and makes sure that the discussion will stay focused on the topic given to get the insightful results at the end. Typically, the moderator asks different types of questions during the interview to gather the most relevant information from the group [16].

Section 3.1 discusses the focus group methodology that was used and details the demographics of the participants. Section 3.2 explains the characteristics and finally, Section 3.3 discusses the relationship between these characteristics.

### 3.1 Study Design

The focus group methodology performed for this research involved a total of nine participants including four intermediate gamers (i.e., playing games at least three to four times a week), two experienced gamers (i.e., playing games for almost 8-10 years), while the remaining were less casual gamers (i.e., they played different game genres at least twice a month). The group consisted of seven male and two female gamers with an age range of 19-27 years and a mean age of 23 years.

A short introductory presentation was given to the participants at the start of the focus group to provide a brief understanding of cloud gaming and how delay can affect gaming QoE. Also, some important findings of the research community such as the precision-deadline model by Claypool were explained [12] to the participants.

Next, all participants played 12 different game scenarios picked from different game genres with 0 ms, 150 ms, and 300 ms delay. These scenarios were related to the various game types, and each scenario was played for 60 seconds. Once a participant played all the scenarios, they wrote post-game notes about the impact of different scenarios in the game as well as their observations about the factors which can make playing the game with delay intolerant. Afterwards, in the second phase of this study, a group discussion was made, moderator tried to lead the discussion but tried to avoid bias the participants. All the participants were given liberty to openly discuss the game characteristics which were identified to potentially influence the delay sensitivity of the game. Thereby, only characteristics which can be visually quantifiable by someone with reasonable gaming knowledge should be considered. In the end, the participants presented a list of nine game characteristics that can be reasons for the difference in delay tolerance among the different scenarios of the games.

### 3.2 Characteristics Definition

The inputs collected from the focus group were compiled to give them understandable definitions. The following nine characteristics with respect to the sensitivity of game scenarios towards delay were driven:

**Temporal Accuracy (TA):**
Temporal Accuracy describes the available time interval for a player to perform a desired interaction. The time interval is strongly dependent on the mechanics and pace of a game scenario. This characteristic was inspired by deadline as defined by Claypool [12].

The characteristic could be quantified by using a 6-point category scale with the labels unlimited, long, moderate, short, extremely short and immediate. An example of a game with an unlimited time interval to perform an interaction is chess without time restrictions. As an example, a game that requires an immediate response is a shooting game in which a player who reacts first (immediately) wins.

### Spatial Accuracy (SA):

The spatial accuracy is the degree of precision required to complete an interaction successfully. Typically, game scenarios in which the player must select (or point at) on object precisely, or in which precise movements are necessary, require a high spatial accuracy. This characteristic often strongly depends on the size of the controlled object (e.g., a character or curser) and the size of objects to interact with (e.g., a target or platform). This characteristic was inspired by precision as defined by Claypool [12].

The characteristic should be quantified by using a 4-point category scale with the labels no required accuracy, low required accuracy, moderately required accuracy, and high required accuracy. Here, an example for a game scenario not requiring any precision is a Flipper game, as only the timing of the paddle movement is important. Using a sniper weapon in a shooting game scenario, would be an example for a game scenario which requires a high spatial accuracy.

### Predictability (PR):

Predictability describes if a player is able to estimate the upcoming events in the game. This can for example relate to positions of objects (spatial) or time points of events (temporal).

The characteristic should be quantified by using a 4-point category scale with the labels nothing to predict, easy to predict, difficult to predict and not predictable. An example for a game that does not have anything to predict in the action level is a card playing game. And a game that is difficult to predict is a shooting game against humans.

### The number of Input Directions (NID):

The number of possible input directions in a game scenario is known as Degree of Freedom (DoF). DoF consists of translations (back and forward, left and right, up and down) as well as rotations (vertical axis and height) for one or multiple input devices/elements.

The characteristic should be quantified by using a 5-point category scale with the labels 1, 2, 3, 4, and more than 4. An example of a game scenario with DoF of 1, a jumping game where a player must simply jump (up) can be considered. In a jumping game where, in addition to jumping, a player can also go forward and backward, the DoF is 3. In a shooting game where a player can go in 4 different directions with a keyboard and 4 different directions with a mouse, the DoF is 8.

### Consequences (CQ):

How strong a delayed interaction influences a user's experience, often depends on the negative consequences due to failing to perform the desired action. Such consequences could be the loss of progress, points, and rewards. This characteristic is similar to the characteristic impact defined by Claypool [17].

The characteristic should be quantified by using a 3-point category scale with the labels of low, medium, and high. In a scenario without negative consequences of an incorrect interaction or in which the game status is only worse, but it is still realistic to win (e.g., chess or racing game), the consequences are low. If an incorrect interaction would result in a situation in which the game is difficult to win afterward, the consequences are medium. Lastly, if an incorrect interaction immediately results in losing the game (e.g. jump and run in which a collision with an object directly leads to losing), the negative consequences are high.

### Importance of Actions (IoA):

The Importance of Actions describes how much each input of a game scenario can change its outcome. There are games in which every input of the user can significantly change the outcome of the game, and others which are more tolerant towards errors.

The characteristic should be quantified by using a 3-point category scale using the labels low, medium and high. Exploring a game world or looking at a map would be a scenario with a low importance of actions. Selecting units in a strategy game or shooting with a weapon which has a high shooting rate (e.g., a minigun), can be considered as moderately important actions. A jump and run game scenario or shooting at a target with a weapon which has a low shooting rate (e.g., a sniper) would mainly consist of highly important actions.

### The number of Required Actions (NRA):

The number of required actions and with that also the number of inputs a player performs in a certain time frame may influence the perception of a network delay. The characteristic could also be described as the minimum actions per minute (APM) to play a game scenario. It is assumed that a higher number of required actions will lead to more user inputs and thus, more situations in which a player can perceive a delay. The number of objects to react to or the pace of the game can influence the number of required actions.

The characteristic should be quantified by using a 3-point category scale using the labels low, moderate and high. A game scenario with a low number of required actions would be puzzle games, in which every few seconds (more than 2 seconds) an action is performed. A game scenario that requires an interaction every 1-2 seconds should be considered as moderate in terms of the number of required actions. Lastly, a scenario in which more than one interaction is required every second should be rated as a scenario with a high number of required actions.

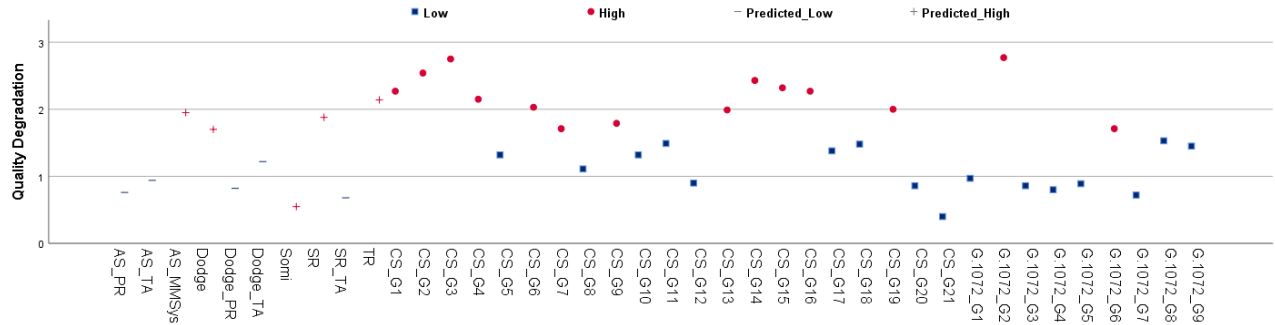

Figure 1. The degradation of IQ on 0ms and 200ms in 40 games (30 training and 10 test). The training games are classified using k-means into two clusters, red points are high sensitive games and blue points are games having have low sensitivity to delay.

**Feedback Frequency (FF):**
Feedback frequency means how often the game gives visual, auditive, or haptic feedback to the player. A player's input frequency may influence the degree to which he perceives a possible delay.

The characteristic should be quantified by using a 3-point category scale with the labels rarely, sometimes and very often. An example of a game scenario that rarely gives feedback to the user would be a game where the users are always holding the same key and an example of a game that very often gives feedback to the user would be a game that the users are required to keep moving a mouse cursor.

**Type of Input (ToI):**
The type of input describes the temporal aspects of player inputs on a spectrum of discrete to continuous. In some games, players are continuously giving input, for example in a shooting game where players are always moving their mouse. Some games have discrete inputs meaning that players interact using pressing a button, for example, a jumping game where players must jump using pressing a key. In games with Quasi–Continuous inputs players interact with the game using holding a key or constantly pressing a key.

The characteristic should be quantified by using a 5-point category using the labels Quasi-Continuous, Quasi-Continuous and discrete, Only Discrete input type, Only Continuous input type and Continuous and Discrete.

### 3.3 Characteristics Grouping

Based on the subjective experiment that is detailed in Section 4, a Principal Component Analysis (PCA) was done on the characteristics to get more insight into the characteristics and their relationships. It turns out that these characteristics can be separated into three factors where F1 are the characteristics that are mostly related to the game design itself, F2 summarizes the characteristics related to the game scoring system, and F3 is linked to the characteristics that are related to the user interaction and user inputs. Figure 2 shows these three factors.

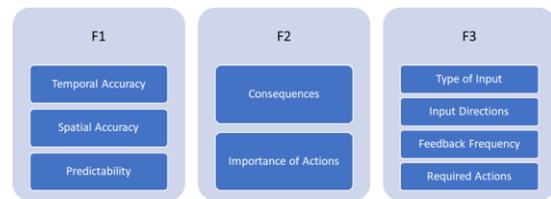

Figure 2. Overview of PCA of the nine identified characteristics.

### 4 Characteristics Quantifications

In order to quantify the identified characteristics, another study was conducted. The aim of this quantification is to build a dataset of game characteristic ratings for the development of the decision tree and to investigate the consistency between the user's ratings to evaluate the reliability of the expert quantification method.

### 4.1 Demographic Information and Questionnaire

To quantify the characteristics, 14 expert gamers including 13 males and one female were invited to take part in the study. Subjects were aged between 20 and 33 years old (M = 23.79 years) and well experienced gamers with the average gaming experience of 3.6 (on a 5-point discrete scale ranging from novice to expert) and well aware of the impact of delay on gaming experience indicated by an average self-judgement of delay awareness of 4.2 on a discrete 5-point agreement scale ranging from 'strongly disagree' to 'strongly agree'.

Participants at first were given clear instruction, and each of the identified characteristic was defined clearly using figures and examples. After the instruction, they all had to pass a training phase, in which a game video was shown to the users. Participants were asked in addition to the rating of the characteristics to write a reason for each of their decision. The questionnaire instructions and the training phase can be seen in the link on the link below: https://github.com/blindsubmission2020/MMVE20

Afterwards, 30 gaming videos were shown to the participants and they were asked to rate all the characteristics. These videos were recorded with a duration of 30 seconds on the same scenario as the details about the dataset is given in the section 5. In addition to the video, a short-written description of the game rules and objectives was given to the participants. These videos were randomized using a Latin Square design to prevent any ordering effect.

## 4.2 Validation of Scales and Definitions

In this section the reliability of the subjective ratings will be evaluated to investigate if the identified characteristics are well defined and the scales were clear for the experts. A measure is highly reliable when it produces similar results between different users. The absolute agreement between users is shown in Table 1. It is computed using a two-way random Inter-Class Correlation (ICC) for all the characteristics.

The results, summarized in Table 1, indicate for TA, SA, NID and NRA there was an excellent degree of agreement between the users as indicated by an ICC > 0.9. For the characteristic CQ and PR a good level, ICC > 0.8, and for FF, a fair level of agreement, ICC > 0.7, was reached. However, for the characteristic IoA the agreement between the raters was at a poor level which shows that the scales and definition for this characteristic need to be modified.

**Table 1: Performance of the scales and definitions in terms of the agreement between the experienced participants.**

| Characteristic | ICC | CI | F | p |
|---|---|---|---|---|
| TA | 0.94 | [.91, .97] | 21.66 | .001 |
| SA | 0.92 | [.88, .96] | 14.98 | .001 |
| PR | 0.81 | [.70, .89] | 6.89 | .001 |
| NID | 0.95 | [.93, .97] | 30.01 | .001 |
| CQ | 0.88 | [.81, .93] | 9.64 | .001 |
| IoA | 0.59 | [.37, .76] | 2.86 | .001 |
| NRA | 0.90 | [.84, .94] | 12.33 | .001 |
| FF | 0.72 | [.56, .84] | 4.83 | .001 |
| ToI | 0.89 | [.82, .94] | 12.07 | .001 |

## 5 The Decision Tree

In this section, the decision tree is proposed upon the characteristics that are explained in Section II and the ratings of the IQ collected from a mixed dataset containing 30 games. In these two datasets 570 subjects were participated. These 30 games include nine games from the interactive study of ITU-T recommendation G.1072 [18], in addition to the Crowdsourcing games in [19] as where using nine different open-source games and modifying different characteristics of a total of 21 games was created. Both of these two datasets were shared privately by the authors upon our request.

In the Crowdsourcing study, 375 participants including 146 females, 225 males and 4 others participated. The range of age for these participants was between 18-60 years with a mean age of 36 years. In the interactive study of G.1072, 195 participants including 84 females and 111 males participated. The range of age for these participants was 19 to 42 years with a mean age of 28.82 years.

All of these games were played once with 200ms delay and once with 0ms delay. The 200ms delay was chosen based on [3], as it was shown to be most diverging to indicate low and high sensitive games towards delay. IQ is the average of responsiveness and the controllability of the test condition [20].

These games were clustered using K-means based on the drop of IQ. The optimum number of clusters turns out to be two clusters (silhouette value = 0.77). We refer them as the low (C1) and high (C2) sensitive class. Figure 1. shows these clusters for all 30 games based on the drop of IQ from 0ms to 200ms.

Afterwards, to map the game characteristics to the clusters, we calculated a decision tree based on the assigned game characteristic values. The visual representation of the decision tree is shown in Figure 3. In the end, two classes of sensitivity were defined based on four characteristics, ToI, NID, PR and TA. The performance of classification in terms of accuracy, precision recall and F1 score is reported in Table 2. In addition, the confusion matrix is reported in Table 3. The tree achieved an overall accuracy of 86.6 %.

On the left side the statistics of Table 3, presents the original class assigned and the upper row describes the predicted class in the decision tree. The analysis indicates that the performance of Cluster 1 is 93.7% and for cluster 2 reached 78.5%. In total, there were 4 mismatches between the decision tree and the subjective results. From our point of view, two of these four differences are caused by too optimistic subjective ratings for first person shooter games in the dataset, which are not in line with other studies investigating these games. This might be due to the expertise of the participants or selection of the scenario. If we do not consider these two errors, the decision tree accuracy reaches 93%.

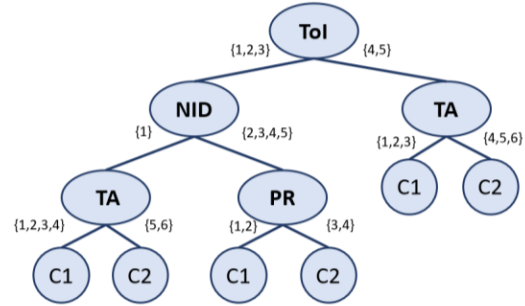

**Figure 3. Delay sensitivity decision tree depending upon different gaming characteristics.**

**Table 2: Performance of delay sensitivity decision tree.**

| Accuracy | Precision | Recall | Specificity | F1 |
|---|---|---|---|---|
| 0.86 | 0.83 | 0.93 | 0.78 | 0.88 |

**Table 3: Confusion matrix of delay sensitivity decision tree.**

| Total | | Predicted | | |
|---|---|---|---|---|
| | | C1 | C2 | Total |
| Actual | C1 | 15 | 1 | 16 |
| | C2 | 3 | 11 | 14 |
| | Total Achieved | 18 | 12 | 30 |

### 5.1 Validation

In addition to the training set the validation was done using 10 games form previous datasets [20], [21]. These games including simple shooting games, racing games and platform jumping game. These 10 games were played under 0 and 200ms delay at least by a minimum of 25 participants and they were asked to rate the IQ. Afterwards the drop of IQ from 0ms to 200ms was calculated. The datapoints from the test set are also included in Figure 1. Similar to the training set, videos of these 10 games were recorded and shown and quantified by experts. Since the consistency between the raters were shown to be high, we invited only three experienced gamers (playing games for almost 8-10 years) for quantification. Participants were asked to rate the four characteristics of the decision tree. Based on the expert ratings, the decision tree could predict 9 out of 10 games correctly. The only mismatch was belonging to the game Somi which is a 2D shooting game in which a gun is fired permanently and thus, no precise timing is required. Table 4 shows the performance of the decision tree on the test set in terms of accuracy, precision, recall, specificity and F1 score.

**Table 4: Performance of delay sensitivity decision tree.**

| Accuracy | Precision | Recall | Specificity | F1 |
|---|---|---|---|---|
| 0.9 | 0.8 | 1 | 0.83 | 0.88 |

## 6  Discussion and Conclusion

This paper presents an expert evaluation method to classify games with respect to their delay sensitivity. Using this classification, cloud provider, game developers, and network planners can have a better insight into the game requirement and can deliver a better QoE to their users. The classification should be specifically used in the context of cloud gaming and it is not applicable in online gaming where the lag compensation techniques are dominant. The classification uses some characteristics that are required to be quantified by watching a video of the game. The evaluations show an excellent level of the agreement between the users which indicates with even a low number of experienced gamers these characteristics can be quantified. In future work, we will investigate more on the minimum number of required experts. In addition, some of the characteristics such as APM, ToI and NID can be computed objectively by looking at the users' input as shown in [22].

## 7 Conclusion

This paper presents 9 characteristics which potentially influence the sensitivity of a video game used for cloud gaming services towards delay. These characteristics were defined and an expert evaluation methodology to quantify these characteristics was presented. Afterwards, using a dataset of 30 games, the characteristics were quantified, and the games were mapped based on the degradation (comparing 200 ms and 0 ms delay conditions) of subjective ratings of IQ by means of a decision tree. The results show that the decision tree has an accuracy of 90% on the test set. In addition, the excellent level of agreement between the raters shows that the characteristics and scales can provide reliable results.